\documentclass[aps,pra,reprint,superscriptaddress,longbibliography]{revtex4-2}

\usepackage{amsmath,amsthm,amsfonts}
\usepackage[colorlinks=true,citecolor=blue,linkcolor=blue,urlcolor=blue]{hyperref}

\usepackage{float}
\usepackage{graphicx}
\usepackage{tabularx}
\usepackage[font=small]{caption}
\usepackage{subcaption}
\usepackage{physics}
\usepackage{mhchem}
\usepackage{quantikz}
\usepackage{lipsum}
\usepackage{multirow}
\usepackage{svg}

\newcolumntype{M}[1]{>{\centering\let\newline\\\arraybackslash\hspace{0pt}}m{#1}}
\newcolumntype{Q}{M{0.12\textwidth}}
\newcolumntype{O}{M{0.12\textwidth}}
\newcolumntype{I}{M{0.2\textwidth}}
\newcolumntype{C}{M{0.08\textwidth}|M{0.13\textwidth}|M{0.1\textwidth}}

\begin{document}

\title{Quantum Computation of Frequency-Domain Molecular Response Properties Using a Three-Qubit iToffoli Gate}

\author{Shi-Ning Sun}
\thanks{These authors contributed equally to this work}
\affiliation{Division of Engineering and Applied Science, California Institute of Technology, Pasadena, CA 91125, USA}
\thanks{These authors contributed equally to this work}

\author{Brian Marinelli}
\thanks{These authors contributed equally to this work}
\affiliation{Quantum Nanoelectronics Laboratory, Department of Physics,
University of California at Berkeley, Berkeley, CA 94720, USA}
\affiliation{Computational Research Division, Lawrence Berkeley National Laboratory, Berkeley, California 94720, USA}

\author{Jin Ming Koh}
\affiliation{Division of Physics, Mathematics and Astronomy, California Institute of Technology, Pasadena, California 91125, USA}

\author{Yosep Kim}
\affiliation{Center for Quantum Information, Korea Institute of Science and Technology (KIST), Seoul 02792, Korea}

\author{Long B.~Nguyen}
\affiliation{Quantum Nanoelectronics Laboratory, Department of Physics, University of California at Berkeley, Berkeley, CA 94720, USA}
\affiliation{Computational Research Division, Lawrence Berkeley National Laboratory, Berkeley, California 94720, USA}

\author{Larry Chen}
\affiliation{Quantum Nanoelectronics Laboratory, Department of Physics, University of California at Berkeley, Berkeley, CA 94720, USA}
\affiliation{Computational Research Division, Lawrence Berkeley National Laboratory, Berkeley, California 94720, USA}

\author{John Mark Kreikebaum}
\affiliation{Quantum Nanoelectronics Laboratory, Department of Physics, University of California at Berkeley, Berkeley, CA 94720, USA}
\affiliation{Materials Science Division, Lawrence Berkeley National Laboratory, Berkeley, California 94720, USA}

\author{David I. Santiago}
\affiliation{Quantum Nanoelectronics Laboratory, Department of Physics, University of California at Berkeley, Berkeley, CA 94720, USA}
\affiliation{Computational Research Division, Lawrence Berkeley National Laboratory, Berkeley, California 94720, USA}

\author{Irfan Siddiqi}
\affiliation{Quantum Nanoelectronics Laboratory, Department of Physics, University of California at Berkeley, Berkeley, CA 94720, USA}
\affiliation{Computational Research Division, Lawrence Berkeley National Laboratory, Berkeley, California 94720, USA}
\affiliation{Materials Science Division, Lawrence Berkeley National Laboratory, Berkeley, California 94720, USA}

\author{Austin J.~Minnich}
\email{aminnich@caltech.edu}
\affiliation{Division of Engineering and Applied Science, California Institute of Technology, Pasadena, CA 91125, USA}

\date{\today}

\begin{abstract}
The quantum computation of molecular response properties on near-term quantum hardware is a topic of significant interest. While computing time-domain response properties is in principle straightforward due to the natural ability of quantum computers to simulate unitary time evolution, circuit depth limitations restrict the maximum time that can be simulated and hence the extraction of frequency-domain properties. Computing properties directly in the frequency domain is therefore desirable, but the circuits require large depth when the typical hardware gate set consisting of single- and two-qubit gates is used. Here, we report the experimental quantum computation of the response properties of diatomic molecules directly in the frequency domain using a three-qubit iToffoli gate, enabling a reduction in circuit depth by a factor of two. We show that the molecular properties obtained with the iToffoli gate exhibit comparable or better agreement with theory than those obtained with the native CZ gates. Our work is among the first demonstrations of the practical usage of a native multi-qubit gate in quantum simulation, with diverse potential applications to the simulation of quantum many-body systems on near-term digital quantum computers.

\end{abstract}

\maketitle

\section*{Introduction}

A primary goal of emerging quantum computing technologies is to enable the simulation of quantum many-body systems that are challenging for classical computers \cite{feynman_1982,lloyd_1996,georgescu_2014}. Early experimental demonstrations of quantum simulation algorithms have focused on computing ground- and excited-state energies of small molecules \cite{peruzzo_2014,omalley_2016,kandala_2017,colless_2018} or few-site spin \cite{kandala_2017} and fermionic models \cite{barends_2015}. More recently, the scale of quantum simulation experiments has increased in terms of numbers of qubits, diversity of gate sets, and complexity of algorithms, as manifested in simulation of models based on real molecules and materials \cite{google_2020,tazhigulov_2022,stanisic_2022}, various phases of matter such as thermal \cite{motta_2019,francis_2021}, topological \cite{satzinger_2021,tan_2021}
and many-body localized states \cite{mi_2022,karamlou_2022}, as well as holographic quantum simulation using quantum tensor networks
\cite{niu_2022,chertkov_2022,gibbs_2022}. As quantum advantages in random sampling have been established on quantum hardware \cite{google_2019,wu_2021}, focus has turned to the experimental demonstration of quantum advantages in problems of physical significance \cite{daley_2022}.

\begin{figure*}
\centering{
\phantomsubcaption\label{fig:diatomic_molecules}
\phantomsubcaption\label{fig:diagonal_circuit}
\phantomsubcaption\label{fig:off_diagonal_circuit}
\includegraphics[width=0.87\textwidth]{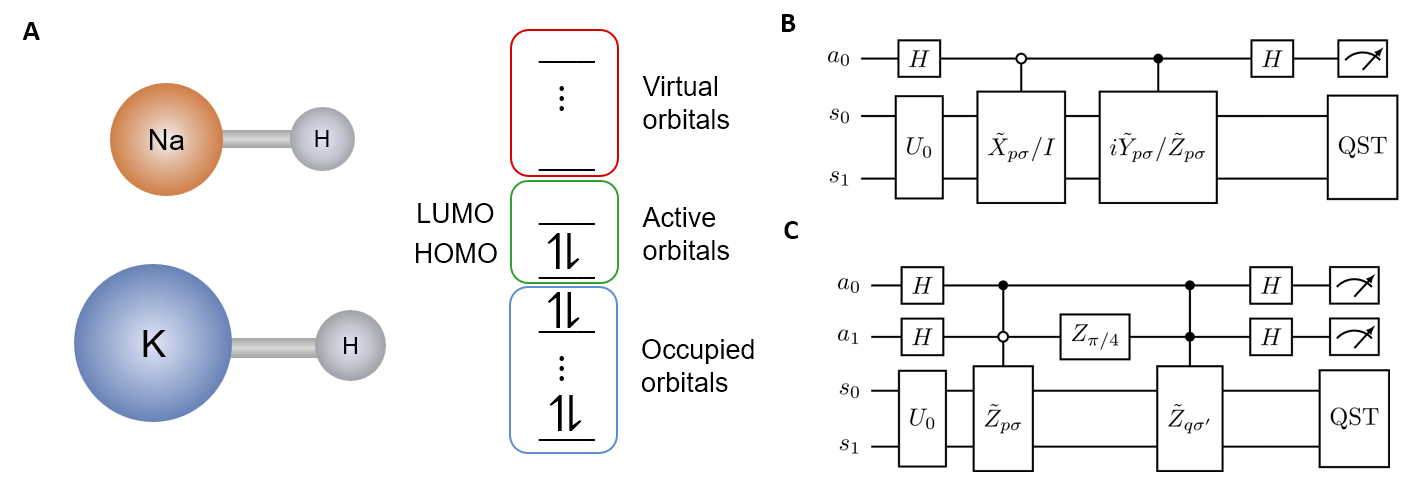}}
\caption{\textbf{Schematic of the diatomic molecules and LCU circuits for computing transition amplitudes.} ($\textbf{A}$) Schematic of the diatomic molecules NaH and KH. The active space consists of only the highest occupied molecular orbital (HOMO) and the lowest unoccupied molecular orbital (LUMO). (\textbf{B}) The circuits to calculate diagonal transition amplitudes, where $a_0$ is the ancilla qubit and $s_0$ and $s_1$ are the system qubits. For the spectral functions the target unitaries are $\tilde{X}_{p\sigma}$ and $i\tilde{Y}_{p\sigma}$, while for the response function the target unitaries are $I$ and $\tilde{Z}_{p\sigma}$. (\textbf{C}) The circuit to calculate off-diagonal transition amplitudes in the response functions, where $a_0$ and $a_1$ are the ancilla qubits, and $s_0$ and $s_1$ are the system qubits. The double-controlled-$\tilde{Z}$ gates are decomposed with either iToffoli gates or CZ gates. The double-controlled identity gates that would complete the standard two-ancilla LCU circuit is not shown.}
\label{fig:algo_and_circuits}
\end{figure*}

For applications in chemistry and physics, the calculation of the response properties of molecules and materials is of substantial interest \cite{cao_2019,mcardle_2020,bauer_2020}. Investigating response properties in the electronic structure theory framework involves calculating quantities such as the one-particle Green's function \cite{von_nissen_1984} and density-density response functions \cite{ullrich_2019}, which provide insight into interpreting experimental spectroscopic measurements \cite{damascelli_2003}. Response properties of molecules and materials can be determined either in time domain or in frequency domain. Due to the natural ability of quantum computers to simulate time evolution \cite{feynman_1982,lloyd_1996}, near-term algorithms to compute time-domain response properties have been carried out on quantum hardware \cite{chiesa_2019,francis_2020,sun_2021}. However, computing the frequency-domain response from the time-domain response using the typical gate set available on hardware requires a time duration that  exceeds the circuit depth limitations of near-term quantum computers.

An alternative approach to determine these response properties is by computing them directly in the frequency domain. Frequency-domain algorithms generally involve obtaining the ground- and excited-state energies as well as the transition amplitudes between the ground state and the excited states. Although there are established methods to obtain ground- and excited-state energies on quantum computers \cite{mcclean_2017,nakanishi_2019,parrish_2019,higgott_2019,jones_2019}, calculating transition amplitudes is less straightforward. Various schemes including variational quantum simulation \cite{endo_2020,chen_2021}, quantum subspace expansion \cite{jamet_2022} and quantum linear algebra  \cite{cai_2020} to determine frequency-domain response properties have been proposed. However, the accuracy of variational methods depends on the quality of the ansatz, quantum subspace expansion is susceptible to numerical instabilities from basis linear dependence, and quantum linear algebra is out of reach for near-term quantum hardware. Recently, a non-variational scheme amenable to near-term hardware implementation has been proposed \cite{kosugi_2020_construction,kosugi_2020_linear}. This scheme constructs the electron-added and electron-removed states simultaneously by exploiting the probabilistic nature of the linear combination of unitaries (LCU) algorithm \cite{childs_2012}. Nevertheless, this LCU-based algorithm has not yet been demonstrated on quantum hardware due to the lack of efficient implementations of the multi-qubit gates.

In this work, we experimentally demonstrate the calculation of frequency-domain response properties of diatomic molecules using a recently reported high-fidelity three-qubit iToffoli gate \cite{kim_2022} to implement the LCU circuits on a superconducting quantum processor. The multi-controlled gates present in the original LCU algorithm are decomposed either with iToffoli gates or with CZ gates. We calculate transition amplitudes between the ground state and the $N$-electron or $(N\pm 1)$-electron states of NaH and KH molecules restricted to the highest occupied and lowest unoccupied molecular orbitals (HOMO and LUMO). The transition amplitudes are then used to construct spectral functions and density-density response functions of the diatomic molecules. We apply error mitigation techniques including randomized compiling (RC) \cite{wallman_2016,hashim_2021_randomized} during circuit construction, and McWeeny purification \cite{mcweeny_1960} during postprocessing, both of which result in marked improvement of the experimental observables. The observables obtained from the reduced-depth circuits with iToffoli decomposition show comparable or better agreement with theory compared to the observables from circuits with CZ decomposition, despite incomplete Pauli twirling in the randomized compiling procedure applied to the iToffoli gate. Our results pave the way for the application of multi-qubit gates for quantum chemistry and related quantum simulation applications on near-term quantum hardware.

\section*{Results}
\subsection*{Quantum Algorithm for Transition Amplitudes of Diatomic Molecules}

We consider the HOMO-LUMO models of the diatomic molecules NaH and KH as shown in Fig.~\ref{fig:diatomic_molecules} (see Methods). Such molecular models with reduced active space have been used in benchmarking quantum chemistry methods on quantum computers \cite{yeter-aydeniz_2021}. The HOMO-LUMO model generates two spatial orbitals or equivalently four spin orbitals, which correspond to four qubits after Jordan-Wigner transformation \cite{jordan_1928}. To reduce quantum resources, we exploit the number symmetry in each spin sector to reduce the number of qubits from four to two using a qubit-tapering technique \cite{bravyi_2017} (details given in Supplementary Sec.~\ref{sec:z2_symmetry_transforms}).

The observables we aim to determine are the spectral function and density-density response function. Suppose that the molecular Hamiltonian with reduced active space has ground state $\ket{\Psi_0}$ with energy $E_0$, and $(N\pm 1)$-electron eigenstates $\ket{\Psi_\lambda^{N\pm 1}}$ with energies $E^{N\pm 1}_\lambda$. Let $\hat{a}_{p\sigma}^\dagger$ and $\hat{a}_{p\sigma}$ be the creation and annihilation operators on orbital $p$ with spin $\sigma$, respectively. The one-particle Green's function has the expression \cite{von_nissen_1984}:

\begin{align}
G_{pq}(\omega) &= \sum_{\lambda\sigma} \frac{\langle\Psi_0 |\hat{a}_{p\sigma}| \Psi_{\lambda}^{N+1} \rangle \langle \Psi_{\lambda}^{N+1}|\hat{a}^\dagger_{q\sigma}| \Psi_0\rangle}{\omega + E_0 - E_{\lambda}^{N+1} + i\eta} \notag \\
&+ \sum_{\lambda\sigma} \frac{\langle\Psi_0 |\hat{a}_{q\sigma}^\dagger| \Psi_{\lambda}^{N-1} \rangle \langle \Psi_{\lambda}^{N-1}|\hat{a}_{p\sigma}| \Psi_0\rangle}{\omega - E_0 + E_{\lambda}^{N-1} + i\eta}
\label{eq:G}
\end{align}
where $\omega$ is the frequency and $\eta$ is a small broadening factor. The spectral function is related to the Green's function by $A(\omega) = -\pi^{-1}\Im\, \Tr\, G(\omega)$.

For the density-density response function, we consider the charge-neutral $N$-electron excited states $\ket{\Psi_\lambda^N}$ with energies $E_\lambda^N$ and the number operator $\hat{n}_{p\sigma}$ on the orbital $p$ with spin $\sigma$. The density-density response function has the expression \cite{ullrich_2019}:
\begin{align}
R_{pq}(\omega) = &\sum_{\lambda} \frac{\sum_{\sigma\sigma'}\langle\Psi_0|\hat{n}_{p\sigma}|\Psi_\lambda^N\rangle \langle \Psi_\lambda^N|\hat{n}_{q\sigma'}|\Psi_0\rangle}{\omega + E_0 - E_\lambda^{N} + i\eta}.
\label{eq:R}
\end{align}

The operators $\hat{a}_{p\sigma}, \hat{a}_{p\sigma}^\dagger$ and $\hat{n}_{p\sigma}$ are not unitary, but they can be written as linear combination of unitary operators as
\begin{align}
\hat{a}_{p\sigma} &= (\bar{X}_{p\sigma} - i\bar{Y}_{p\sigma})/2,\\
\hat{a}_{p\sigma}^\dagger &= (\bar{X}_{p\sigma} + i\bar{Y}_{p\sigma})/2,\\
\hat{n}_{p\sigma} &= (I - Z_{p\sigma})/2,
\end{align}
where $I$ is the identity operator, $Z_{p\sigma}$ is the Pauli $Z$ operator on orbital $p$ with spin $\sigma$, and $\bar{X}_{p\sigma}$ and $\bar{Y}_{p\sigma}$ are the Jordan-Wigner transformed Pauli $X$ and $Y$ operators on orbital $p$ with spin $\sigma$ with a string of $Z$ operators included to account for the anticommutation relation \cite{jordan_1928}. The Pauli strings $\bar{X}_{p\sigma}, \bar{Y}_{p\sigma}$ and $Z_{p\sigma}$ undergo the same transformation and qubit tapering process as the Hamiltonian (details given in Supplementary Sec.~\ref{sec:z2_symmetry_transforms}). Except for the identity operator which does not change under the transformation, we label the transformed $\bar{X}_{p\sigma}, \bar{Y}_{p\sigma}, Z_{p\sigma}$ as $\tilde{X}_{p\sigma}, \tilde{Y}_{p\sigma}$ and $\tilde{Z}_{p\sigma}$.

The LCU circuits to calculate diagonal and off-diagonal transition amplitudes are given in Figs.~\ref{fig:diagonal_circuit} and \ref{fig:off_diagonal_circuit}. Each circuit has two system qubits $s_0$ and $s_1$, and one ancilla qubit $a_0$ or two ancilla qubits $a_0$ and $a_1$. The unitary $U_0$ prepares the ground state $\ket{\Psi_0}$ on the system qubits from the all-zero initial state. The circuit diagrams reflect that the spectral function involving operators $\tilde{X}_{p\sigma}$ and $\tilde{Y}_{p\sigma}$ only requires diagonal transition amplitudes, while the density-density response function involving operators $I$ and $\tilde{Z}_{p\sigma}$ requires both the diagonal and off-diagonal transition amplitudes. Although the original algorithm \cite{kosugi_2020_construction,kosugi_2020_construction} proposed performing quantum phase estimation on the system qubits, due to quantum resource constraints we instead apply quantum state tomography \cite{christandl_2012} to the system qubits while measuring the ancilla qubits in the $Z$ basis.

In the diagonal circuits, we obtain the (unnormalized) system-qubit states $\frac{1}{2}(\tilde{X}_{p\sigma} \pm i\tilde{Y}_{p\sigma})\ket{\Psi_0}$ or $\frac{1}{2}(I \pm \tilde{Z}_{p\sigma})\ket{\Psi_0}$ with probabilities $p_{\pm}$, where the probabilities are specified by the ancilla measurement outcome as $p_+ = p_{a_0 = 0}$ and $p_- = p_{a_0 = 1}$; in the off-diagonal circuits, we obtain the (unnormalized) system-qubit states $\frac{1}{4}[(I - \tilde{Z}_{p\sigma}) \pm e^{i\pi/4}(I - \tilde{Z}_{q\sigma'})]\ket{\Psi_0}$ with probabilities $p_\pm$, where $p_+ = p_{(a_0,a_1) = (1, 0)}$ and $p_- = p_{(a_0, a_1) = (1, 1)}$. We take the overlap of the tomographed system-qubit states with the exact eigenstates, which are then postprocessed according to Eq.~18 in Ref.~\cite{kosugi_2020_construction} or Eq.~25 in Ref.~\cite{kosugi_2020_linear} to yield the transition amplitudes (see Supplementary Sec.~\ref{sec:transition_amplitudes}). The transition amplitudes are then used to construct the spectral function and density-density response function according to Eqs.~\ref{eq:G} and \ref{eq:R}.

In the following sections, for simplicity, we will denote the diagonal circuit that applies the operator $\hat{a}^{(\dagger)}_{p\sigma}$ or $\hat{n}_{p\sigma}$ as the $p\sigma$-circuit, and the off-diagonal circuit that applies the operators $\hat{n}_{p\sigma}$ and $\hat{n}_{q\sigma'}$ as the $(p\sigma, q\sigma')$-circuit.

\begin{figure}
\centering{
\phantomsubcaption\label{fig:doubled_controlled_gate}
\phantomsubcaption\label{fig:itoffoli_decomposition}
\phantomsubcaption\label{fig:cz_decomposition}
\includegraphics[width=0.36\textwidth]{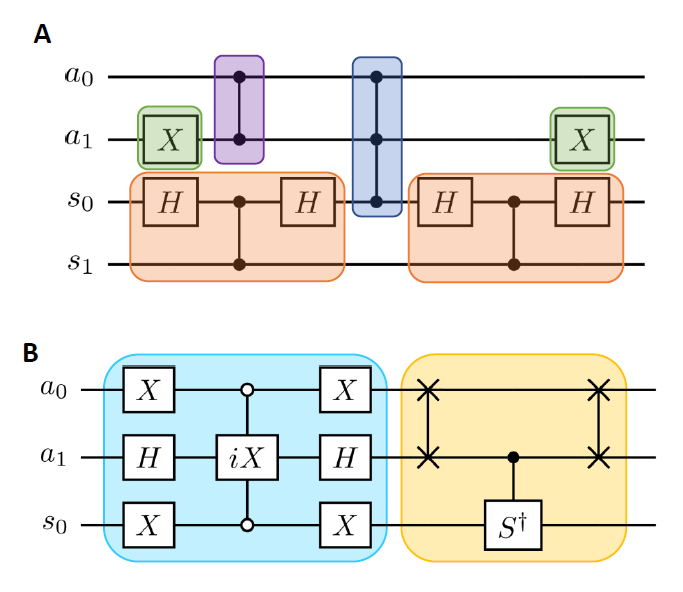}}
\caption{\textbf{Decomposition of the double-controlled composite gates in the LCU circuits.} (\textbf{A}) Example of the decomposition of a double-controlled $-ZZ$ gate into CCZ (blue) along with other single- and two-qubit gates. The $X$ gates (green) are used to adjust the control states; the CZ gate on $a_0$ and $a_1$ (purple) is used to adjust the overall multiplicative factor of the Pauli string, which is $-1$ in this case; the equivalent CNOT gates (orange) are used to extend the Pauli string as in Ref.~\cite{whitfield_2011}. (\textbf{B}) Decomposition of the CCZ gates with the iToffoli gate, which includes a CC-iZ part (light blue) and an equivalent long-range CS dagger part (yellow). The SWAP gates are simplified in the transpilation stage or further decomposed with CZ gates according to Ref.~\cite{hashim_2021_optimized}.}
\label{fig:ccz_decomposition}
\end{figure}

\subsection*{iToffoli versus CZ Decomposition in LCU Circuits}

The transformed and tapered operators are two-qubit Pauli strings with multiplicative factors of $\pm 1$ or $\pm i$. To apply the single- or double-controlled gates, we follow the standard multi-qubit Pauli gate decomposition \cite{whitfield_2011} with the base gate as CZ or CCZ and use CNOT gates consisting of native CZ gates dressed by Hadamard gates to extend the weights of the Pauli strings. The multiplicative factor $-1$ or $\pm i$ can be applied as a single-qubit phase gate on the ancilla in the diagonal circuits, or as the native CS, CS$^\dagger$ or CZ on the two ancillae in the off-diagonal circuits. Additionally, $X$ gates are wrapped around the ancilla qubits controlled on the 0 state. Figure \ref{fig:doubled_controlled_gate} shows how a double-controlled gate with ancilla $a_0$ controlled on 1, ancilla $a_1$ controlled on 0, and a target operator $-ZZ$ is applied on the device.

\begin{figure}[t]
\centering{
\phantomsubcaption\label{fig:nah_spectral_function}
\phantomsubcaption\label{fig:kh_spectral_function}
\includegraphics[width=0.48\textwidth]{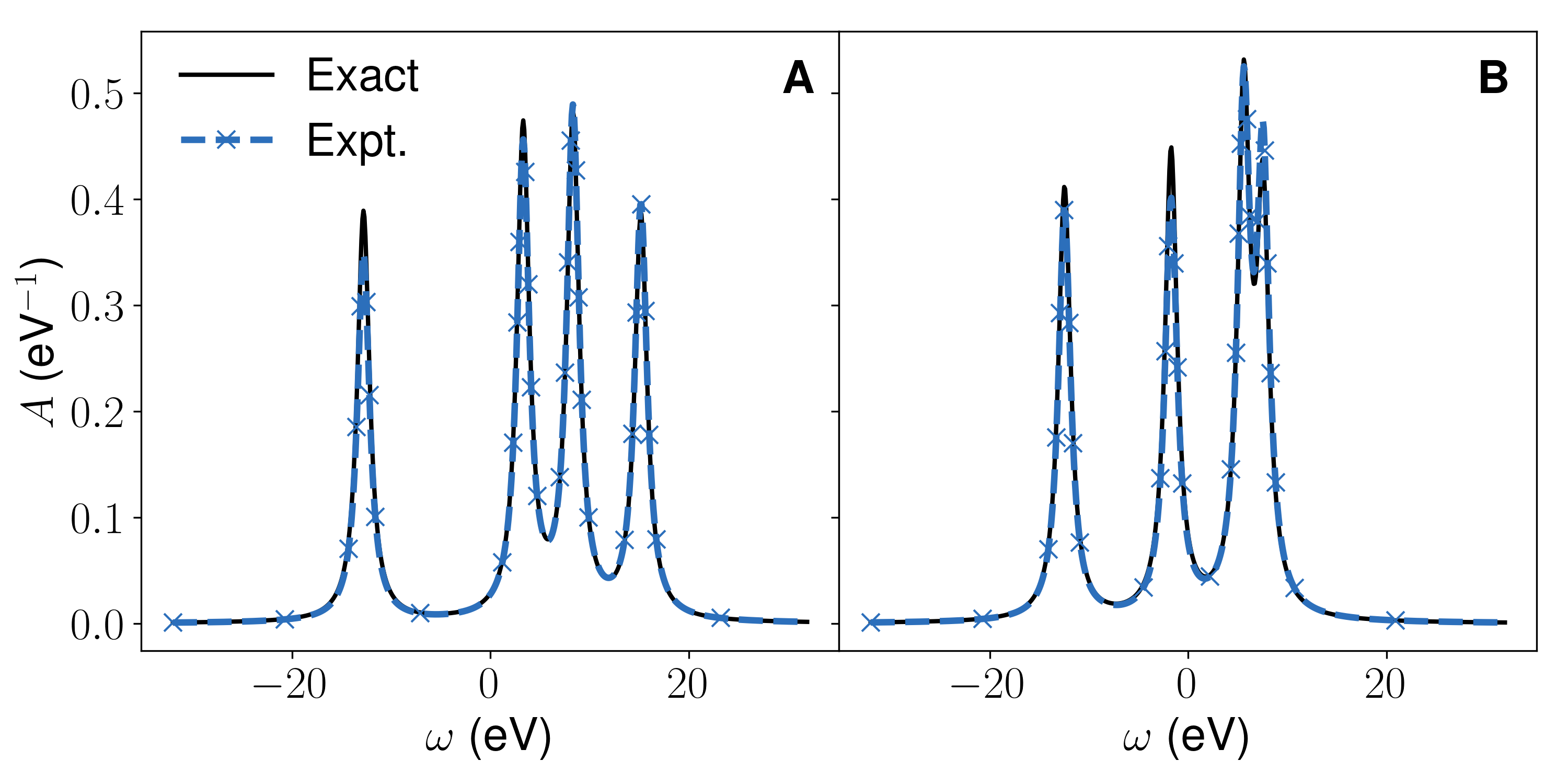}}
\caption{{\color{black}\textbf{Spectral function of diatomic molecules.} Spectral function of (\textbf{A})  NaH, (\textbf{B}) KH. The circuits to obtain the spectral function are shallow three-qubit circuits that do not require the iToffoli gates. The experimental spectral functions are in quantitative agreement with the exact ones, with maximum peak height deviation of 10.6\%.}}
\label{fig:spectral_function}
\end{figure}

\begin{figure}[b]
\centering{
\phantomsubcaption\label{fig:nah_fidelity_vs_depth}
\phantomsubcaption\label{fig:kh_fidelity_vs_depth}
\includegraphics[width=0.4\textwidth]{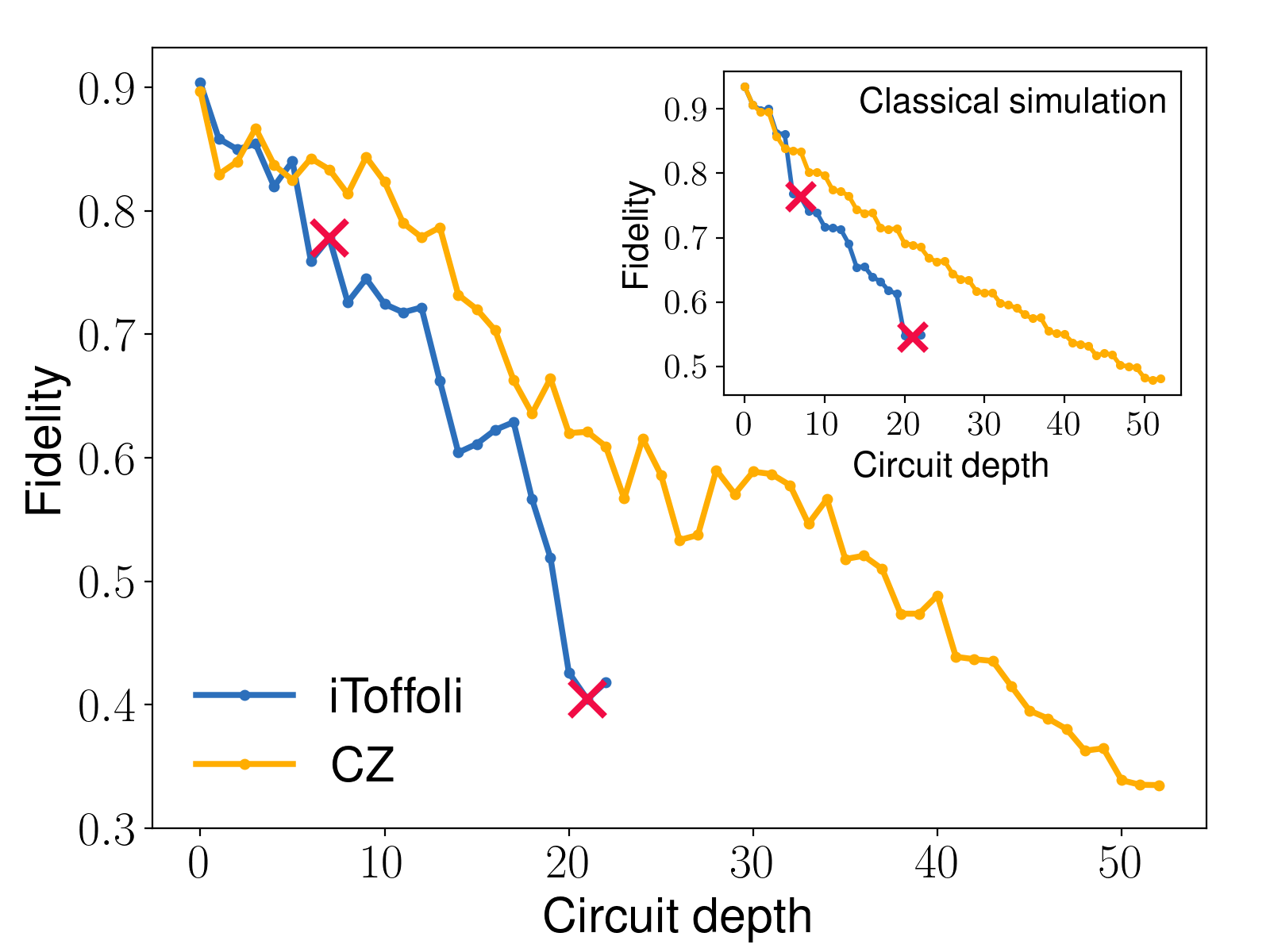}}
\caption{\textbf{Fidelity versus circuit depth of the $(0\uparrow, 0\downarrow)$-circuit for NaH.} Fidelity for the iToffoli decomposition (blue) and the CZ decomposition (yellow). The locations of the iToffoli gates are marked by red crosses. The CZ decomposition results in lower overall fidelity compared to iToffoli decomposition due to higher circuit depth. The inset is the corresponding data from noisy simulation and shows a similar trend. All results in this figure are raw experimental or simulated data without any error mitigation.}
\label{fig:fidelity_trajectory}
\end{figure}

We decompose the CCZ gate either with the three-qubit iToffoli gate as shown in Fig.~\ref{fig:itoffoli_decomposition} or with the native CZ gates. The iToffoli decomposition starts with a double-controlled $i$Z component, followed by a long-range CS$^\dagger$ gate to cancel the phase factor $i$. The SWAP gates in the long-range CS$^\dagger$ part of the circuit are further simplified in the transpilation stage or decomposed into three CZ gates and additional single-qubit gates according to a recent work on the same quantum device \cite{hashim_2021_optimized}. For the CZ decomposition of CCZ, we use a topology-aware quantum circuit synthesis package \cite{younis_2021} to obtain the optimal decomposition as eight CZs under linear qubit connectivity, as opposed to the six-CZ decomposition that requires all-to-all qubit connectivity \cite{shende_2009}. 

The spectral function only requires the four diagonal circuits $0\uparrow, 0\downarrow, 1\uparrow, 1\downarrow$. The density-density response function requires four diagonal circuits $0\uparrow, 0\downarrow, 1\uparrow, 1\downarrow$ and six off-diagonal circuits $(0\uparrow, 0\downarrow), (0\uparrow, 1\uparrow), (0\uparrow, 1\downarrow), (0\downarrow, 1\uparrow), (0\downarrow, 1\downarrow), (1\uparrow, 1\downarrow)$. We use the same transpilation procedure to optimize the circuits constructed from iToffoli decomposition and CZ decomposition (details given in Methods). The diagonal circuits after transpilation are relatively shallow circuits with maximum circuit depth (excluding virtual $Z$ gates) of 19, maximum two-qubit gate count of 7 and no iToffoli gates. In the off-diagonal circuits, the circuit depths range from 24 to 29 for iToffoli decomposition and from 54 to 59 for CZ decomposition. As for the two- and multi-qubit gate counts, each iToffoli-decomposed circuit contains two iToffoli gates and 9 to 12 native two-qubit gates, while each CZ-decomposed circuit contains 19 to 21 native two-qubit gates. Hence the iToffoli decomposition results in about half the circuit depth and half the number of two-qubit gates compared to the CZ decomposition.

\subsection*{Spectral Function and Response Function on Quantum Hardware}

The spectral function of NaH and KH are shown in Fig.~\ref{fig:spectral_function}. The density matrices are obtained from quantum state tomography and postprocessed with McWeeny purification. Randomized compiling is not employed for these results. The experimental spectral functions show very good agreement with the exact ones, with maximum peak height deviation of 10.6\%.

\begin{figure}
\centering{
\phantomsubcaption\label{fig:fidelity_matrix_norc_raw}
\phantomsubcaption\label{fig:fidelity_matrix_rc_raw}
\phantomsubcaption\label{fig:fidelity_matrix_norc_pur}
\phantomsubcaption\label{fig:fidelity_matrix_rc_pur}
\includegraphics[width=0.49\textwidth]{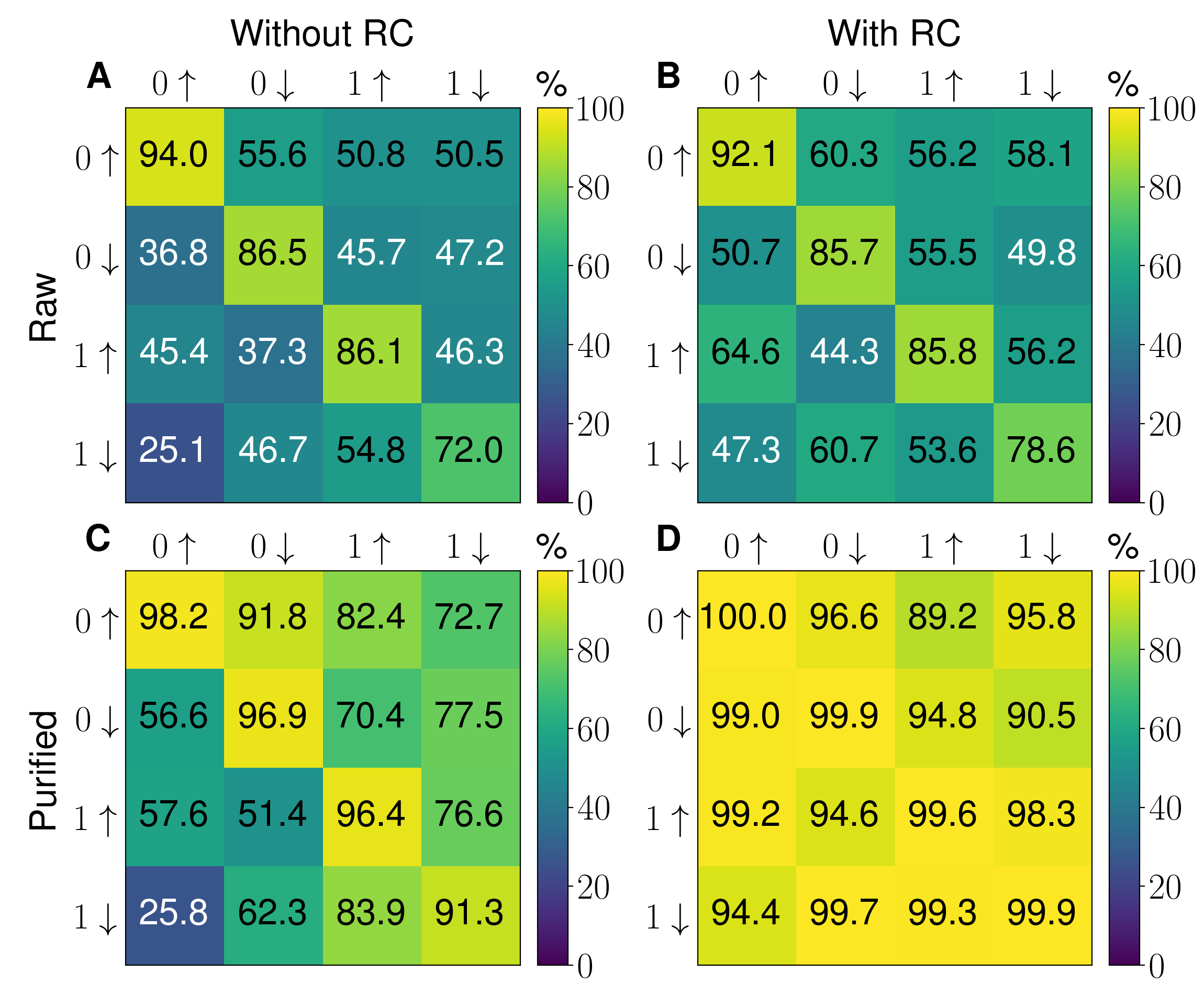}}
\caption{\textbf{System-qubit state fidelities in the response function calculation of NaH.}  (\textbf{A} to \textbf{B}) Fidelities between the raw experimental and exact system-qubit density matrices without (A) and with RC (B). The diagonal elements correspond to system-qubit density matrices in the diagonal circuits after taking the ancilla state $a_0 = 1$, and the off-diagonal elements correspond to the system-qubit density matrices in the off-diagonal circuits after taking the ancilla states either as $(a_0, a_1) = (1, 0)$ (upper diagonal) or as $(a_0, a_1) = (1, 1)$ (lower diagonal). (\textbf{C} to \textbf{D}) Fidelities between the purified experimental and exact system-qubit density matrices without (C) and with RC (D). Layout of the tiles are the same as in panels (A) and (B). Without RC, purification raises the average off-diagonal fidelity from 45.2\% to 67.4\%, but with both RC and purification the average off-diagonal fidelity increases to 96.0\%.
}
\label{fig:fidelity_matrix}
\end{figure}

\begin{figure*}
\centering{
\phantomsubcaption\label{fig:chi00_norc}
\phantomsubcaption\label{fig:chi00_rc}
\phantomsubcaption\label{fig:chi01_norc}
\phantomsubcaption\label{fig:chi01_rc}
\includegraphics[width=0.62\textwidth]{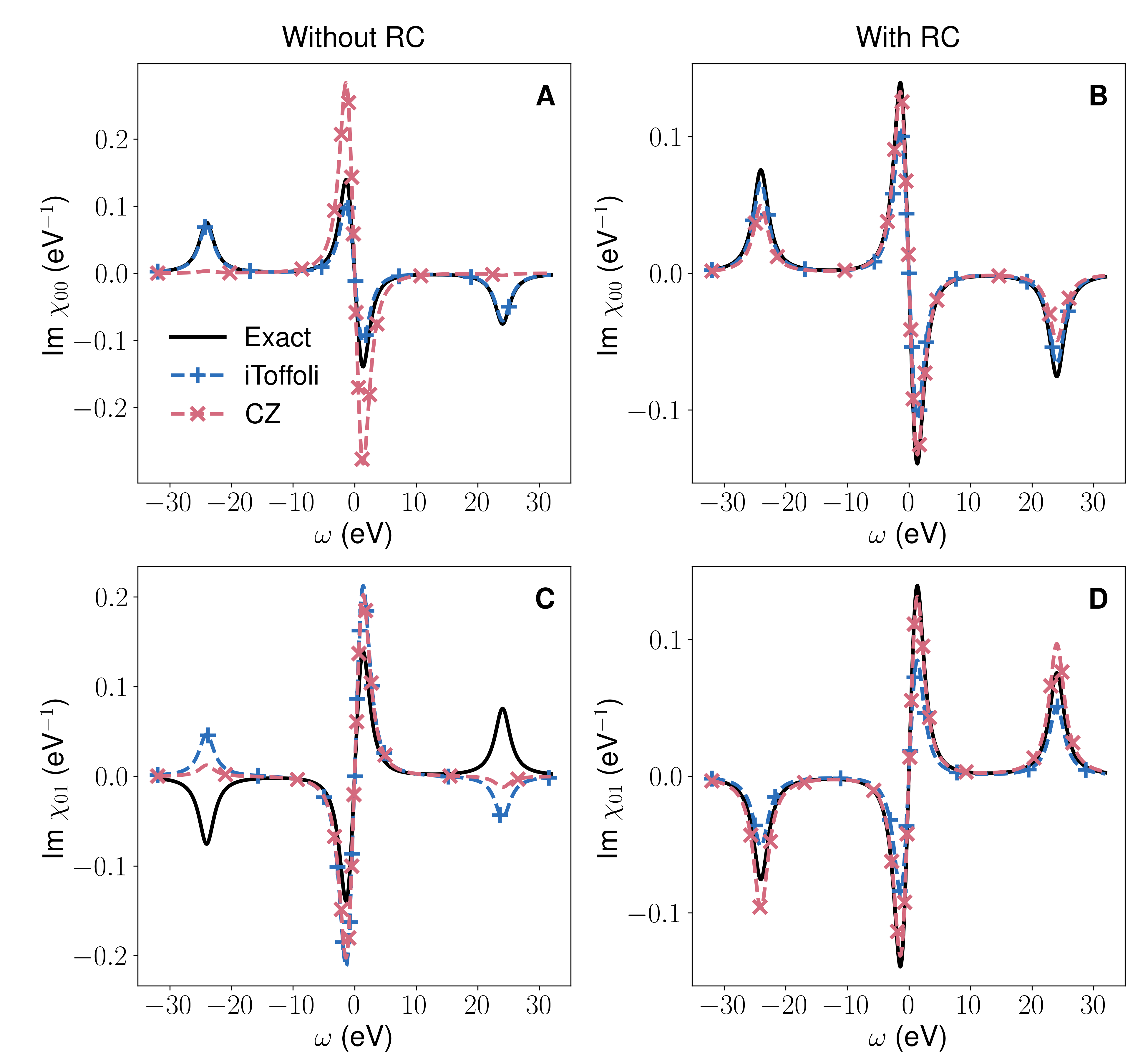}}
\caption{\textbf{Density-density response function of NaH}. (A) $\Im\,\chi_{00}$ without RC. (B) $\Im\,\chi_{00}$ with RC. (C) $\Im\,\chi_{01}$ without RC. (D) $\Im\,\chi_{01}$ with RC.  All experimental results are postprocessed with McWeeny purification on the system-qubit states after restricting to the ancilla bitstring subspace. Without RC, the iToffoli decomposition yields qualitatively improved results compared to the CZ decomposition, as quantified by the root-mean-squared (RMS) error of 0.0086 eV$^{-1}$ (iToffoli) compared to 0.0372 eV$^{-1}$ (CZ) in $\Im\,\chi_{00}$; with RC, the two decompositions exhibit comparable RMS error values of 0.0081 eV$^{-1}$ (iToffoli) vs 0.0073 eV$^{-1}$ (CZ) in $\Im\,\chi_{00}$.}
\label{fig:response_function}
\end{figure*}

We next turn to the density-density response functions, which are more challenging to determine than the spectral functions because these observables require the deeper off-diagonal circuits containing three-qubit iToffoli gates. We begin by considering a specific off-diagonal circuit needed for the density-density response function, the $(0\uparrow, 0\downarrow)$-circuit. To understand the influence of the iToffoli gate on the accuracy of the executed circuit, we compute the fidelity of the whole qubit register obtained by quantum state tomography versus circuit depth. The same quantity was computed for a circuit using only CZ gates to decompose the double-controlled gates. The results are shown in Fig.~\ref{fig:fidelity_trajectory}. Although the iToffoli decomposition shows a steeper decrease in fidelity compared to the CZ decomposition, the fidelity at the end of the circuit is higher due to lower circuit depth. The noisy simulation in the inset of Fig.~\ref{fig:fidelity_trajectory} shows a similar trend. The iToffoli gate reported in Ref.~\cite{kim_2022} does not consider spectator errors on neighboring qubits, which are cancelled out in the gate calibration in this work (details given in Supplementary Sec.~\ref{sec:gate_calibration}). The cycle benchmarking fidelity of the iToffoli gate accounting for the spectator qubit is 96.1\%, lower than the single-qubit gate fidelities which are above 99.5\% and the two-qubit gate fidelities which are between 97.6\% and 98.8\%, which may explain the steeper decay in fidelity with circuit depth in the iToffoli circuit compared to the CZ circuit.

Next, we examine the fidelity of the final state in each iToffoli-decomposed circuit used in the calculation of response functions. Figure \ref{fig:fidelity_matrix} shows the system-qubit state fidelities on each response function circuit for NaH, where McWeeny purification is applied to the system-qubit density matrix after restricting the full density matrix to each ancilla bitstring sector. Comparing the values in Fig.~\ref{fig:fidelity_matrix_norc_raw} with those in Fig.~\ref{fig:fidelity_matrix_rc_raw}, we can see that RC itself only results in a moderate improvement in the fidelities, with the average diagonal fidelities changing from 84.6\% to 85.5\% and average off-diagonal fidelities changing from 45.2\% to 54.8\%. However, the results between Fig.~\ref{fig:fidelity_matrix_rc_raw} and Fig.~\ref{fig:fidelity_matrix_rc_pur} show that RC combined with purification yield an average diagonal fidelity of 99.9\% and an average off-diagonal fidelity of 96.0\%, even though purification without RC only leads to a limited improvement in the average diagonal fidelity from 85.6\% to 95.7\%, and in the average off-diagonal fidelity from 45.2\% to 67.4\% in Figs.~\ref{fig:fidelity_matrix_norc_raw} and \ref{fig:fidelity_matrix_norc_pur}.

We now show the density-density response functions $\chi_{00}$ and $\chi_{01}$ of NaH in Fig.~\ref{fig:response_function}. Here $\chi_{00}$ is obtained from two diagonal circuits $0\uparrow, 0\downarrow$ and one off-diagonal circuit $(0\uparrow, 0\downarrow)$, while $\chi_{01}$ is obtained from four off-diagonal circuits $(0\uparrow, 1\uparrow), (0\uparrow, 1\downarrow),(0\downarrow, 1\uparrow),(0\downarrow, 1\downarrow)$. All experimental results are postprocessed with purification after constraining the ancilla qubits to each bitstring subspace.

Overall, the iToffoli decomposition yields comparable or better results compared to the CZ decomposition. Without RC, the iToffoli decomposition results in visibly better agreement with the exact response functions in Figs.~\ref{fig:chi00_norc} and \ref{fig:chi01_norc}, where the root-mean-squared (RMS) errors between the experimental and exact response functions are 0.0086 eV$^{-1}$ and 0.0372 eV$^{-1}$ for the iToffoli decomposition and 0.0387 eV$^{-1}$ and 0.0278 eV$^{-1}$ for the CZ decomposition for $\chi_{00}$ and $\chi_{01}$, respectively. With RC, the two decompositions exhibit comparable RMS errors, which are 0.0081 eV$^{-1}$ and 0.0143 eV$^{-1}$ for the iToffoli decomposition, and are 0.0073 eV$^{-1}$ and 0.0061 eV$^{-1}$ for the CZ decomposition for $\chi_{00}$ and $\chi_{01}$, respectively, as shown in Figs.~\ref{fig:chi00_rc} and \ref{fig:chi01_rc}.

In both $\chi_{00}$ and $\chi_{01}$, RC can lead to substantial improvements in the experimental data for both decompositions, which is reflected in the heights of the symmetric peaks at $\pm 1.4$ eV and $\pm 24.0$ eV. For $\chi_{00}$ without RC and using the CZ decomposition, the peak at $\pm 1.4$ eV is twice the height of the exact peak and the peak at $\pm 24$ eV is not present in the experimental observable in Fig.~\ref{fig:chi00_norc}. With RC, both features are present, with the deviations of the peak heights being 34.8\% and 4.7\%, respectively, as in Fig.~\ref{fig:chi00_rc}. For $\chi_{00}$ with the iToffoli decomposition, the peak height deviations change from 6.1\% and 26.6\% without RC in Fig.~\ref{fig:chi00_norc} to 11.8\% and 24.0\% with RC in Fig.~\ref{fig:chi00_rc}. In this case, the results before and after RC are comparable since the experimental observable without RC is already in good agreement with the exact observable. For  $\chi_{01}$, we see a more marked improvement due to RC compared to the corresponding results for $\chi_{00}$. Without RC, both iToffoli and CZ decompositions produce the wrong sign on the peak at $\pm 24$ eV in Fig.~\ref{fig:chi01_norc}, but with RC the signs are correctly predicted. The peak height deviations can then be computed as 39.2\% and 32.2\% for the CZ decompositions and 5.7\% and 28.2\% for the iToffoli decomposition, as indicated in Fig.~\ref{fig:chi01_rc}. The corresponding results of system-qubit state fidelities and density-density response functions for KH are given in Supplementary Sec.~\ref{sec:addtional_data_for_kh}, which follow a similar trend as NaH.



Since the iToffoli gate is non-Clifford, our implementation of RC results in incomplete Pauli twirling compared to applying RC to the CZ-decomposed circuits (see Supplementary Sec.~\ref{sec:randomzied_compiling}). The incompleteness of RC on the iToffoli-decomposed circuits may explain why the two decompositions have similar RMS errors with RC despite the initial advantage for the iToffoli decomposition without RC due to its lower circuit depth.

\section*{Discussion}

We have carried out an LCU-based algorithm to compute the spectral functions and density-density response functions of diatomic molecules from the transition amplitudes determined on a superconducting quantum processor. Using a native high-fidelity iToffoli gate \cite{kim_2022} has enabled the required circuit depth to be reduced by around a factor of two. These resulting circuits produced better agreement with the exact results compared to the circuits constructed only from CZ, CS, and CS$^\dagger$ gates when RC is not employed. We also developed an RC protocol for the non-Clifford iToffoli gate, and have shown that without complete Pauli twirling on the iToffoli gate, the circuits constructed from iToffoli gates gave comparable results as the circuits constructed only from CZ, CS, and CS$^\dagger$ gates with RC. 


The quality of the computed quantities was greatly improved by the use of several error mitigation techniques. Specifically, our results highlight the significance of RC \cite{wallman_2016,hashim_2021_randomized} combined with McWeeny purification \cite{mcweeny_1960} for quantum simulation. McWeeny purification has been widely used in quantum chemistry \cite{goedecker_1999} and started to be exploited in quantum computing for constraining the purity of the output state \cite{google_2020,ville_2022}. Our results have showed that RC or McWeeny purification individually only improves the experimental results to a limited extent, as observed in the change of the average off-diagonal fidelities from 45.8\% to 54.2\% with only RC, and to 67.4\% with only purification in Fig.~\ref{fig:fidelity_matrix}. However, the combination of RC and purification results in a significant improvement in the quality of the results with the system-qubit state fidelities being 96.0\% on average. Moreover, previous works applied purification to the whole qubit register, but we have showed here that the purification scheme can be applied when there is purity constraint on a subset of qubits. Additionally, our work is the first to apply RC to the non-Clifford iToffoli gate. As more native non-Clifford two-qubit and multi-qubit gates become available, our findings may guide future application of RC to non-Clifford gates.

Our work is also among the first to demonstrate the practical use of a native multi-qubit gate in quantum simulation via an LCU-based algorithm. LCU as a general algorithmic framework is not limited to determining transition amplitudes in frequency-domain response properties, but has broader applications in areas such as solving linear systems \cite{childs_2017}, simulating non-Hermitian dynamics \cite{wen_2019}, and preparing quantum Gibbs states \cite{chowdhury_2017}. Besides the LCU algorithm, quantum algorithms such as Shor's algorithm \cite{shor_1997} and Grover's search algorithm \cite{grover_1997} can benefit considerably from native three-qubit gates with reduction in circuit depths and gate counts. Quantum algorithm design and implementation thus far have been mostly restricted to single- and two-qubit gates due to their ease of implementation and demonstrated high fidelity. Meanwhile, early implementations of three-qubit gates \cite{mariantoni_2011,fedorov_2012,reed_2012} were generally slower and more prone to leakage and decoherence compared to the iToffoli gate employed here due to populating higher levels outside the qubit computational space. However, more recent implementations of three-qubit gates \cite{kim_2022,hill_2021,galda_2021} have begun to address these challenges yielding fidelities approaching those achieved with two-qubit gates. Further, they have been carried out on quantum devices with tens of qubits, suggesting their utility for larger-scale quantum devices. As such native multi-qubit gates become more prevalent, our work paves the way for using them as native gate components in future quantum algorithm design and implementation.

\section*{Methods}

\subsection*{Quantum Circuit Construction}

The molecular orbitals used in this work are in STO-3G basis with molecular integrals determined from PySCF \cite{pyscf_2020}. We use OpenFermion \cite{openfermion_2020} to map the second-quantized Hamiltonian to qubit operators. The ground-state preparation gate on the system qubits are determined classically by constructing a unitary that maps the all-zero initial states to the ground state and then decomposed into three CZ gates and single-qubit gates using the $KAK$ decomposition \cite{vatan_2004}. The LCU circuits are then constructed by applying the gates shown in Figs.~\ref{fig:diagonal_circuit} and \ref{fig:off_diagonal_circuit}, where the SWAP gates are decomposed according to the scheme in Ref.~\cite{hashim_2021_optimized} and the circuits are transpiled by the functions {\tt MergeInteractions}, {\tt MergeSingleQubitGates} and {\tt DropEmptyMoments} in Cirq \cite{cirq_2022}. The transition amplitudes are then combined with the classically determined ground- and excited-state energies to calculate the spectral functions and response functions (See Supplementary Sec.~\ref{sec:transition_amplitudes}).

\subsection*{Quantum Device}

The quantum device used in this work is a superconducting quantum processor with eight transmon qubits. The algorithm is performed on a four-qubit subset of the device with linear connectivity. Single-qubit gates are performed with resonant microwave pulses. Multiplexed dispersive readout allows for simultaneous state discrimination on all four qubits. CZ gates between all nearest neighbors are performed according to the method in Ref.~\cite{mitchell_2021}. The same method allows for a native CS gate on one pair and a native CS$^{\dagger}$ gate on a different pair, according to the requirements of the algorithm. While single-qubit gates are applied simultaneously, microwave crosstalk requires that all two- and three-qubit gates are applied in separate cycles from each other as well as from any single-qubit gates. TrueQ \cite{trueq} is used for circuit manipulations in the implementation of RC as well as gate benchmarking. Internal software is used to map the circuits to hardware pulses for implementing the native gate set.

\section*{Acknowledgements}

The authors acknowledge A.~Hashim, E.~Younis, Y.~Gao, G.~Li for helpful discussions. S.-N.S. and A.J.M. are supported by the U.S. Department of Energy under Award No. DE-SC0019374. B.M., L.B.N., and L.C. are supported by the Quantum Testbed Program of the Advanced Scientific Computing Research for Basic Energy Sciences program, Office of Science of the U.S. Department of Energy under Contract No. DE-AC02-05CH11231. Y.K. is supported by the KIST research program under grant No. 2E31531. S.-N.S would like to acknowledge the QISE-NET fellowship for promoting the collaboration between Caltech and UC Berkeley in the field of quantum computing.

\section*{Author contributions}

S.-N.S. and A.J.M. conceptualized the project. S.-N.S., B.M., and J.M.Koh contributed to the codebase. B.M., L.B.N., L.C., and Y.K. performed the device calibration and hardware runs of the circuits. J.M.Kreikebaum fabricated the device. S.-N.S. and B.M. analyzed the data. All authors contributed to discussion of the manuscript.

\bibliography{main}

\clearpage
\onecolumngrid
\pagebreak

\begin{center}
\textbf{\large Supplemental Materials: Quantum Computation of Frequency-Domain Molecular Response Properties Using a Three-Qubit iToffoli Gate}
\end{center}
\setcounter{equation}{0}
\setcounter{figure}{0}
\setcounter{table}{0}
\setcounter{page}{1}
\renewcommand{\thefigure}{S\arabic{figure}}
\makeatletter

\section{$\mathbb{Z}_2$ symmetry transformations on operators and qubit states} \label{sec:z2_symmetry_transforms}

In this section, we describe the $\mathbb{Z}_2$ symmetry transformations applied to the qubit operators and ancilla qubit subspaces in the linear combination of unitaries (LCU) algorithm. This transformation converts the four-qubit operators and qubit subspaces into two-qubit ones, which are used in constructing the circuits for hardware runs.

There are four spin orbitals in the diatomic molecules we study in this work. We order them as $0\uparrow, 0\downarrow, 1\uparrow, 1\downarrow$ from left to right, corresponding to the qubit indices 0, 1, 2, 3, in the Pauli strings and qubit state bitstrings. The HOMO-LUMO molecular Hamiltonians of NaH and KH have the same number of up-spin and down-spin electrons. After Jordan-Wigner transformation, the parity of up-spin and down-spin electrons correspond to the qubit operators $ZIZI$ and $IZIZ$, which are $\mathbb{Z}_2$ symmetries of the Hamiltonian. The mean-field ground state $\ket{\Phi_0} = \ket{1100}$ has the expectation values $\langle ZIZI \rangle = \langle IZIZ \rangle = -1$, as does the exact ground state $\ket{\Psi_0}$. 

We can define three types of states in our calculation based on $\mathbb{Z}_2$ symmetries: the ``up-spin'' states are the states that have symmetries $\langle ZIZI\rangle = 1, \langle IZIZ\rangle = -1$, which are obtained by applying $\hat{a}_{p\uparrow}$ or $\hat{a}_{p\uparrow}^\dagger$ on the ground state; the ``down-spin'' states are the states that have symmetries $\langle ZIZI\rangle = -1, \langle IZIZ\rangle = 1$, which are obtained by applying $\hat{a}_{p\downarrow}$ or $\hat{a}_{p\downarrow}^\dagger$ on the ground state; the ``spin-balanced'' states are the states that have the symmetries $\langle ZIZI\rangle = -1, \langle IZIZ\rangle = -1$, which are obtained by applying the number operators on the ground state and have the same symmetries as the ground state. Note that the up-spin and down-spin states here are defined from the expectation values of the $\mathbb{Z}_2$ symmetry operators but do not correspond to spin-$z$ components of the corresponding molecular states. For example, the qubit computational state $\ket{0100}$ represents the molecular state with a single electron in the $0\downarrow$ orbital, which has total spin-$z$ expectation value of $-1/2$, but in our definition it is classified as an up-spin state.

For each type of state, we aim to find a $\mathbb{Z}_2$ transformation that generates the minimum number of gates in the circuits that apply the creation or annihilation operators $\hat{a}_{p\sigma}^{(\dagger)}$ or the number operators $\hat{n}_{p\sigma}$. Recall from the main text that in the Jordan-Wigner transformation, the creation or annihilation operators $\hat{a}_{p\sigma}^{(\dagger)}$ have the decomposition $\bar{X}_{p\sigma} \pm i \bar{Y}_{p\sigma}$, where $\bar{X}_{p\sigma}$ and $\bar{Y}_{p\sigma}$ are the Jordan-Wigner transformed Pauli $X$ and $Y$ operators on orbital $p$ with spin $\sigma$, and the number operators $\hat{n}_{p\sigma}$ have the decomposition $I - Z_{p\sigma}$. The transition amplitudes under the $\mathbb{Z}_2$ transformation $U_{\mathbb{Z}_2}$ can be expressed as
\begin{align}
&\langle\Psi_{\lambda}^{N\pm 1}|\hat{a}_{p\sigma}^{(\dagger)}|\Psi_0\rangle = \langle\Psi_{\lambda}^{N\pm 1}|U_{\mathbb{Z}_2}^\dagger U_{\mathbb{Z}_2} \hat{a}_{p\sigma}^{(\dagger)} U_{\mathbb{Z}_2}^\dagger U_{\mathbb{Z}_2}|\Psi_0\rangle = \Big(\langle\Psi_{\lambda}^{N\pm 1}|U_{\mathbb{Z}_2}^\dagger\Big) \Big[ U_{\mathbb{Z}_2} (\bar{X}_{p\sigma} \pm i \bar{Y}_{p\sigma}) U_{\mathbb{Z}_2}^\dagger\Big] \Big(U_{\mathbb{Z}_2}|\Psi_0\rangle\Big),\\
&\langle\Psi_{\lambda}^{N}|\hat{n}_{p\sigma}|\Psi_0\rangle = \langle\Psi_{\lambda}^{N}|U_{\mathbb{Z}_2}^\dagger U_{\mathbb{Z}_2} \hat{n}_{p\sigma} U_{\mathbb{Z}_2}^\dagger U_{\mathbb{Z}_2}|\Psi_0\rangle = \left(\langle\Psi_{\lambda}^{N}|U_{\mathbb{Z}_2}^\dagger\right)\left[ U_{\mathbb{Z}_2}(I - \bar{Z}_{p\sigma})U_{\mathbb{Z}_2}^\dagger\right]\Big( U_{\mathbb{Z}_2}|\Psi_0\rangle\Big),
\end{align}
where the transformed bra state, ket state and operator are grouped in brackets at the end of each equation. On the up-spin states, we use the transformation $U_{\mathbb{Z}_2} = \text{CNOT}(3, 1)\text{CNOT}(2, 0)$; on the down-spin states, we use the transformation $U_{\mathbb{Z}_2} = \text{SWAP}(2, 3)\text{CNOT}(3, 1)\text{CNOT}(2, 0)$, followed by multiplying all the operators by $-1$; on the spin-balanced states, we use the transformation $U_{\mathbb{Z}_2} = \text{CNOT}(2, 3)\text{CNOT}(3, 1)\text{CNOT}(2, 0)$. After the $\mathbb{Z}_2$ transformation, the first two qubits on the operators and the states are then truncated.

As an example, consider the transformation of the Pauli string $\bar{X}_{1\uparrow} = ZZXI$. After we apply the transformation CNOT(3, 1)CNOT(2, 0), the operator becomes $-YZYZ$. To truncate qubit 0 and qubit 1, we need to find the constant factor after the transformed Pauli string acts on the first two qubits of the ground state. The constant factor for $\tilde{X}_{1\uparrow}$ is $\langle 01|YZ|11\rangle = i$, where $\bra{01}$ are the bit values on the first two qubits of the transformed up-spin states $\langle\Psi_\lambda^{N\pm 1}|U_{\mathbb{Z}_2}^\dagger$ and $\ket{11}$ are the bit values on the first two qubits of the transformed ground state $U_{\mathbb{Z}_2}\ket{\Psi_0}$. The factor of $i$ is then combined with the rest of the Pauli string $-YZ$ to give the final truncated form of the Pauli string $\tilde{X}_{1\uparrow} = -iYZ$. The Pauli strings and qubit state bitstrings before and after the $\mathbb{Z}_2$ symmetry transformations and truncations are given in Table \ref{tab:z2_transform}.

\begin{table}[h]
\centering
\begin{tabular}{|O|O|O|O|I|I|}
\hline
          & Transformed Pauli string symbol       & Original Pauli strings & Transformed Pauli strings & Original qubit state bitstrings & Transformed qubit state bitstrings \\ \hline
\multirow{4}{*}{Up-spin} 
& $\tilde{X}_{0\uparrow}$ & $XIII$  & $II$ & $(N+1)$-electron: & $(N+1)$-electron:   \\ 
& $\tilde{Y}_{0\uparrow}$ & $iYIII$ & $ZI$ & 1110, 1011, & 10, 11,\\
& $\tilde{X}_{1\uparrow}$ & $ZZXI$  & $-iYZ$ & $(N-1)$-electron: & $(N-1)$-electron: \\ 
& $\tilde{Y}_{1\uparrow}$ & $iZZYI$ & $-XZ$ & 0100, 0001. & 00, 01. \\ \hline
\multirow{4}{*}{Down-spin}
& $\tilde{X}_{0\downarrow}$ & $ZXII$ & $IZ$ & $(N+1)$-electron: & $(N+1)$-electron: \\
& $\tilde{Y}_{0\downarrow}$ & $iZYII$ & $ZZ$ & 1101, 0111. & 10, 11. \\
& $\tilde{X}_{1\downarrow}$ & $ZZZX$ & $iYI$ & $(N-1)$-electron: & $(N-1)$-electron: \\
& $\tilde{Y}_{1\downarrow}$ & $iZZZY$ & $XI$ & 1000, 0010 & 00, 01. \\ \hline
\multirow{4}{*}{\shortstack{Spin-balanced}} 
& $\tilde{Z}_{0\uparrow}$   & $ZIII$ & $-ZI$ & 1100, & 00, \\
& $\tilde{Z}_{0\downarrow}$ & $IZII$ & $-ZZ$ &  1001, & 01, \\
& $\tilde{Z}_{1\uparrow}$   & $IIZI$ & $ZI$ & 0110, & 10, \\
& $\tilde{Z}_{1\downarrow}$ & $IIIZ$ & $ZZ$ & 0011. & 11. \\ \hline
\end{tabular}
\caption{Pauli strings and qubit state bitstrings under $\mathbb{Z}_2$ transformations and truncations. Each Pauli string is characterized as up-spin or down-spin depending on whether it originates from a creation or annihilation operator on an up-spin or a down-spin orbital, and is characterized as spin-balanced if it originates from a number operator. The qubit state bitstrings are characterized as up-spin, down-spin or spin-balanced by the type of operator that yields the state after applying on the mean-field ground state $|1100\rangle$, which are consistent with the classification based on the expectation values of the $\mathbb{Z}_2$ symmetry operators $ZIZI$ and $IZIZ$ given in the text.}
\label{tab:z2_transform}
\end{table}

\clearpage

\section{Calculation of Transition Amplitudes} \label{sec:transition_amplitudes}

In this section, we give the equations used to calculate transition amplitudes in the spectral function and density-density response function from quantities measured on hardware. We use the same notation as in the main text, where $\ket{\Psi_0}$ is the $N$-electron ground state, $|\Psi_\lambda^{N\pm 1}\rangle$ are the $(N\pm 1)$-electron states, and $|\Psi_\lambda^{N}\rangle$ are the $N$-electron excited states. The transition amplitudes follow the same notation as in Ref.~\cite{kosugi_2020_construction}, where the transition amplitudes from the ground state to the $(N\pm1)$-electron eigenstates in the calculation of spectral functions are

\begin{align}
B_{\lambda,p\sigma,q\sigma'}^{(e)} &= \langle \Psi_{0}| \hat{a}_{p\sigma} | \Psi_{\lambda}^{N+1} \rangle \langle \Psi_{\lambda}^{N+1}| \hat{a}_{q\sigma'}^\dagger | \Psi_0 \rangle,\\
B_{\lambda,p\sigma,q\sigma'}^{(h)} &= \langle \Psi_{0}| \hat{a}_{p\sigma}^\dagger | \Psi_{\lambda}^{N-1} \rangle \langle \Psi_{\lambda}^{N-1}| \hat{a}_{q\sigma'} | \Psi_0 \rangle.
\end{align}

The spectral function only requires the diagonal transition amplitudes. Theoretically the (unnormalized) states after restraining the ancilla to the 0 or 1 state is $\frac{1}{2}(\tilde{X}_{p\sigma}\pm i\tilde{Y}_{p\sigma})\ket{\Psi_0}$. Let the tomographed density matrices corresponding to these states be $\rho_{p\sigma}^{\pm}$. The diagonal transition amplitudes are calculated as the overlap between the $(N\pm1)$-electron eigenstates $|\Psi_\lambda^{N\pm 1}\rangle$ and the tomographed density matrices $\rho_{p\sigma}^{\pm}$:

\begin{align}
B_{\lambda,p\sigma,p\sigma}^{(e)} &= \langle \Psi_\lambda^{N+1} | \rho_{\rho\sigma}^- | \Psi_{\lambda}^{N+1} \rangle, \label{eq:B_e}\\
B_{\lambda,p\sigma,p\sigma}^{(h)} &= \langle \Psi_\lambda^{N-1} | \rho_{\rho\sigma}^+ | \Psi_{\lambda}^{N-1} \rangle \label{eq:B_h}
\end{align}

Similarly, in the calculation of density-density response functions, we follow the notation in Ref.~\cite{kosugi_2020_linear} and define the transition amplitudes from the ground state to $N$-electron eigenstates as

\begin{align}
N_{\lambda,p\sigma,p\sigma} = \langle\Psi_0|\hat{n}_{p\sigma}|\Psi_\lambda^N\rangle\langle\Psi_\lambda^N|\hat{n}_{q\sigma'}|\Psi_0\rangle
\end{align}

In the diagonal circuits, theoretically the (unnormalized) state after restraining the ancilla qubit to the 1 state is $\frac{1}{2}(I - \tilde{Z}_{p\sigma})\ket{\Psi_0}$. Let the corresponding tomographed density matrix obtained from experiments be $\rho_{p\sigma}^{-}$. The diagonal transition amplitudes are calculated by taking the overlap of the $N$-electron eigenstates $|\Psi_\lambda^N\rangle$ and the tomographed density matrix $\rho_{p\sigma}^-$:

\begin{align}
N_{\lambda,p\sigma,p\sigma} = \langle \Psi_\lambda^N |\rho_{p\sigma}^- | \Psi_\lambda^N \rangle
\end{align}

In the off-diagonal circuits, theoretically the (unnormalized) states after restraining the ancilla qubits to $(1, 0)$ or $(1, 1)$ states are $\frac{1}{4}[(I-\tilde{Z}_{p\sigma}) \pm (I-\tilde{Z}_{q\sigma'})]\ket{\Psi_0}$. Let the corresponding tomographed states obtained from experiments be $\rho_{p\sigma,q\sigma'}^{\pm}$. The intermediate transition amplitudes obtained directly from $\ket{\Psi_\lambda^N}$ are defined as

\begin{align}
T_{\lambda,p\sigma,q\sigma'}^{\pm} = \langle\Psi_{\lambda}^N|\rho_{p\sigma,q\sigma'}^{\pm}|\Psi_{\lambda}^N\rangle,
\end{align}

\noindent from which the off-diagonal transition amplitude is determined by Eq.~25 in Ref.~\cite{kosugi_2020_linear} (or equivalently Eq.~18 in Ref.~\cite{kosugi_2020_construction}) as

\begin{align}
N_{\lambda,p\sigma,q\sigma'} = e^{-i\pi/4}(T_{\lambda,p\sigma,q\sigma'}^+ - T_{\lambda,p\sigma,q\sigma'}^-) + e^{i\pi/4}(T_{\lambda,q\sigma',p\sigma}^+ - T_{\lambda,q\sigma',p\sigma}^-). \label{eq:N}
\end{align}

The transition amplitudes $B^{(e)}_{\lambda,p\sigma,p\sigma}, B^{(h)}_{\lambda,p\sigma,p\sigma}$ in Eqs.~\ref{eq:B_e} and \ref{eq:B_h} and $N_{\lambda,p\sigma,q\sigma'}$ in Eq.~\ref{eq:N} are then combined with the ground-state energy $E_0$ as well as the excited-state energies $E^{N\pm 1}_\lambda$ and $E^{N}_\lambda$ to construct the spectral functions in Eq.~\ref{eq:G} and density-density response functions in Eq.~\ref{eq:R}.

\clearpage
\clearpage

\section{Calibration of the iToffoli Gate} \label{sec:gate_calibration}

\begin{figure}[h]
\centering{
\phantomsubcaption\label{fig:meas_protocol}
\phantomsubcaption\label{fig:cond_phase}
\phantomsubcaption\label{fig:spectator_equiv}
\phantomsubcaption\label{fig:cancelled_cond_phase}
\includegraphics[width=0.7\textwidth]{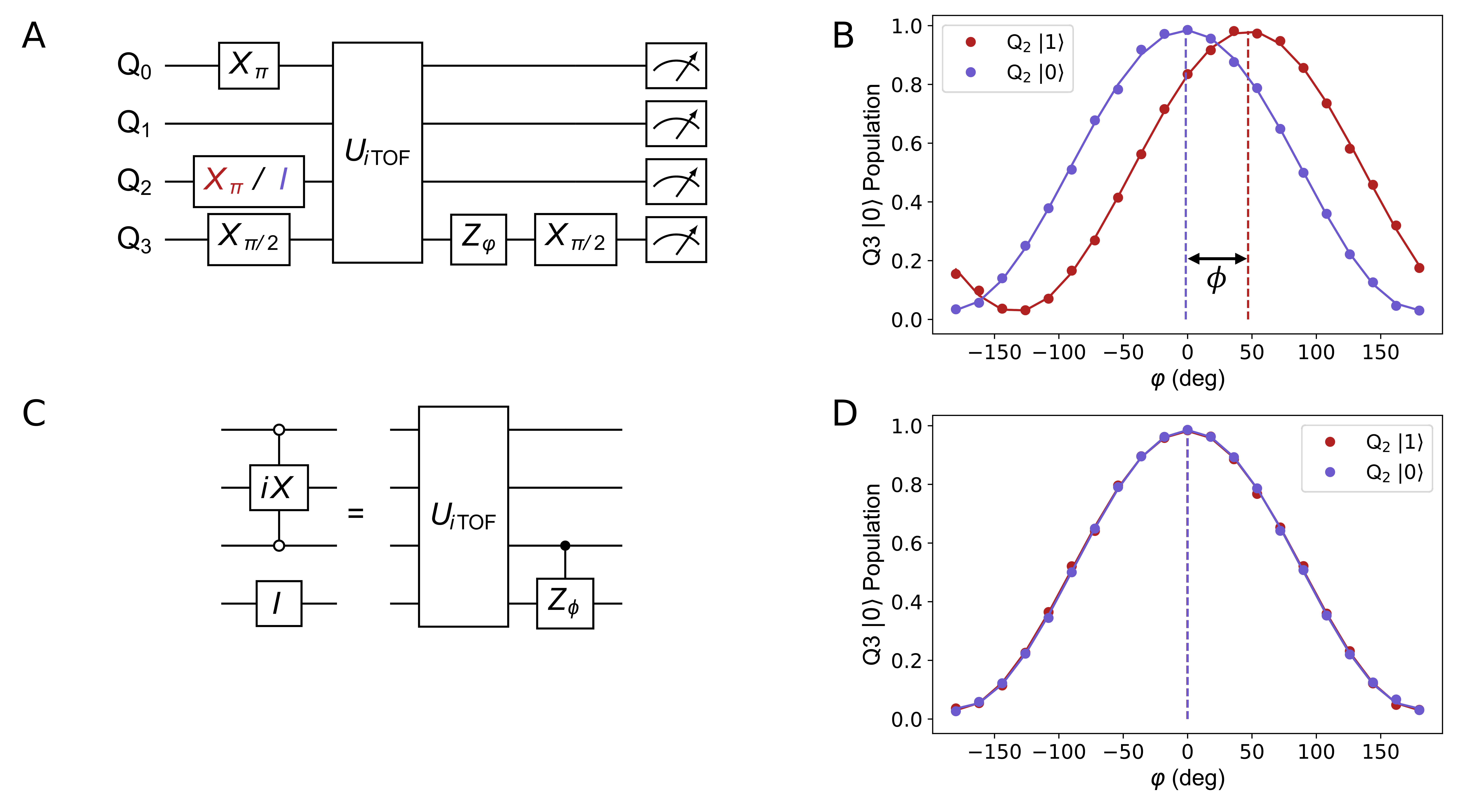}}
\caption{\textbf{Cancellation of spectator error during iToffoli gate.} (A) Ramsey protocol for detecting spurious $ZZ$ error between Q$_{2}$ and spectator Q$_{3}$ during the application of the iToffoli gate. (B) $\ket{0}$ population for Q$_{3}$ after application of the Ramsey sequence in (A) conditional on the state of $Q_{2}$. The relative phase shift between the sinusoidal curves gives the unwanted conditional phase, $\phi$, which must be corrected. (C) A pure iToffoli gate on Q$_{0}$-Q$_{2}$ is achieved by applying the iToffoli drive from Ref.~\cite{kim_2022} following by a $CZ_{\phi}$ gate. (D) Same as (B) except now the $CZ_{\phi}$ correction gate is applied immediately after the iToffoli drive, correcting the unwanted $ZZ$ error.}
\label{fig:spectator_calibration}
\end{figure}

This work employs the recently developed C-$iX$-C iToffoli gate \cite{kim_2022}. In this section we outline the procedure for eliminating spectator errors during the gate application. Since the gate acts on a three qubit subset of the full four qubit subsystem we need to understand and correct the spectator error on the fourth qubit. For concreteness we label qubits $Q_{i}$ for $i=0,\dots,3$ with the iToffoli acting on $Q_{0}$, $Q_{1}$, and $Q_{2}$ with $Q_{3}$ as spectator and denote states in the form $\ket{Q_{0}Q_{1}Q_{2}Q_{3}}$. We run a simple circuit where we prepare the system in either $\ket{100+}$ or $\ket{101+}$ (where $\ket{+}=(\ket{0}+\ket{1})/\sqrt{2}$) and apply the iToffoli gate in a Ramsey-like sequence to determine the $Z$ rotation on $Q_{3}$ conditional on the state of its nearest neighbor, $Q_{2}$ (see Fig.~\ref{fig:meas_protocol} for circuit diagram). We observe an unwanted conditional phase interaction between $Q_{2}$ and $Q_{3}$ with conditional phase, $\varphi = 0.844$ (see Fig.~\ref{fig:cond_phase}). This interaction results from the conditional Stark shift between $Q_{2}$ and $Q_{3}$ when a strong off resonant drive is applied to $Q_{2}$ at the frequency of $Q_{1}$ to implement the iToffoli gate \cite{mitchell_2021}. We can use the same effect to undo the conditional phase by applying simultaneous off-resonant drives to $Q_{2}$ and $Q_{3}$ for a $120~\text{ns}$ period following the iToffoli drive sequence. The full pulse sequence and characterization of the residual conditional rotation with the cancellation applied are shown in Fig.~\ref{fig:spectator_equiv}-\ref{fig:cancelled_cond_phase}. We benchmark the resulting gate implementation using Cycle Benchmarking \cite{erhard_2019}. With this correction, we measure a gate fidelity of $97.8\%$ when isolated to the cycle involving qubits that participate in the iToffoli gate, and a small reduction to $96.6\%$ when including the idling spectator qubit.

\clearpage

\section{Randomized Compiling for Non-Clifford Gates}\label{sec:randomzied_compiling}

In this section we outline a modified version of randomized compiling (RC) \cite{hashim_2021_randomized, wallman_2016} that is applied to the circuits used to compute the observables in the main text. RC is expected to mitigate errors and improve algorithm performance. A broad native entangling gate set is used, consisting of both Clifford gates (CZ) and non-Clifford gates (CS$^{\dagger}$, CS, iToffoli). RC is typically used with hard cycles of $n$-qubit Cliffords in which case the twirling group $\mathcal{T}$ is chosen to be the group of tensor products of $n$ single qubit Paulis. By the definition of Clifford gates, for any Clifford $C$ and twirling gate $T \in \mathcal{T}$ there is some $T^{c} \in \mathcal{T}$ such that $C=TCT^{c}$. RC applied to hard cycles of Clifford gates, $C_{k}$ then proceeds by choosing some $T_{k}$ for each hard cycle $k$, letting $C_{k} \rightarrow T_{k}C_{k}T_{k}^{c}$ and compiling the single qubit Paulis $T_{k}$ and $T_{k}^{c}$ into the easy cycles of single qubit gates before and after, respectively, the Clifford cycle $C_{k}$. 

In order to generalize the method to the non-Clifford gates employed in this work we first find the subsets $\mathcal{T}_{X} \subset \mathcal{T}$ for $X=\mathrm{CS},\mathrm{CS}^{\dagger},\mathrm{iToffoli}$ where for all $T \in \mathcal{T}_{X}$ there is some $T^{c} \in \mathcal{T}_{X}$ such that $X=TXT^{c}$. Then RC proceeds in the same way as above except the twirling gates for hard cycles consisting of gate $X$ are simply chosen from the subset $\mathcal{T}_{X}$ of Pauli strings that stabilize gate $X$. Both the CS and CS$^{\dagger}$ are stabilized by 4 of the possible 16 two-qubit Pauli strings and the iToffoli is stabilized by 8 of the possible 64 three-qubit Pauli strings. For all these non-Clifford gates the twirling and inversion gate are the same, $T=T^{c}$. Results ``with RC'' in the main text involve averaging the experimental bitstring output distributions over 100 equivalent circuit randomizations generated according to the process outlined here, with each circuit measured for 500 shots. These are compared to results ``without RC'' in which the bare circuit is measured for 50000 shots (such that the total number of shots is maintained between the two implementations).

All error processes can be described by a superoperator $\mathcal{E}$ acting on the system density matrix $\rho$. Written in the $n$-qubit Pauli basis this error process matrix is referred to as a Pauli Transfer Matrix (PTM) with diagonal elements giving Pauli fidelities and off-diagonal elements characterizing the unitary (coherent) and non-unitary (incoherent) errors. As discussed in \cite{hashim_2021_randomized, wallman_2016}, applying RC tailors coherent errors into stochastic Pauli noise, which suppresses the off-diagonal elements of the PTM. This holds for the PTM describing errors during the CZ cycles, since these undergo perfect Pauli twirling (in the limit of infinite randomizations). However, in the case of the non-Clifford gate cycles, the twirling is imperfect (since we only twirl over a subset of the $n$-qubit Pauli strings). As a result, some, but not all, of the off-diagonal elements in the corresponding PTMs are suppressed. In other words, not all coherent errors are tailored to stochastic Pauli noise. This imperfect noise tailoring is the main limitation of our approach to generalizing RC to non-Clifford gates. 

We observe a small improvement in the raw state fidelities when using RC but a much larger improvement in the purified state fidelities with respect to the results without RC. The small improvement in the raw fidelities can be explained by the suppression of off-diagonal components of the error process matrix when using RC which lowers the overall error rate slightly. The larger improvement with purification is explained by the fact that RC tailors coherent errors into Stochastic Pauli errors. If the rates of various stochastic Pauli errors are similar, the errors are largely depolarizing and can be corrected by the purification procedure, yielding the high fidelities in Fig.~\ref{fig:fidelity_matrix_rc_pur}. The deviation of the noise from purely depolarizing is responsible (along with the finite number of randomizations) for the remaining infidelity after RC and purification are applied. Conversely, without RC a larger fraction of the error is a coherent over/under rotation of the two-qubit Bloch vector which cannot be corrected by purification.

\clearpage

\section{Additional Data for KH}\label{sec:addtional_data_for_kh}

\begin{figure}[h]
\centering{
\phantomsubcaption\label{fig:fidelity_matrix_norc_raw_kh}
\phantomsubcaption\label{fig:fidelity_matrix_rc_raw_kh}
\phantomsubcaption\label{fig:fidelity_matrix_norc_pur_kh}
\phantomsubcaption\label{fig:fidelity_matrix_rc_pur_kh}
\includegraphics[width=0.49\textwidth]{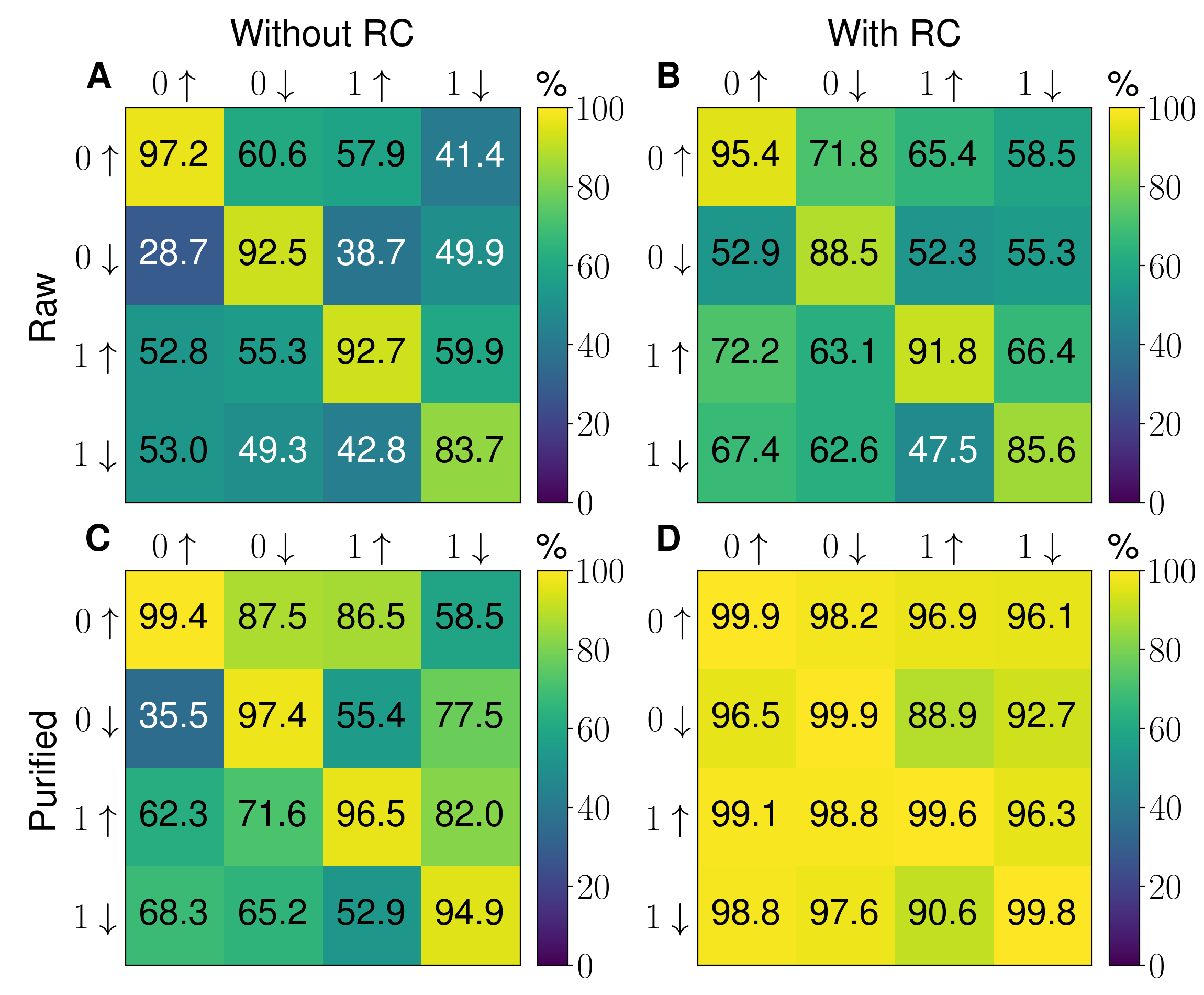}}
\caption{\textbf{System-qubit state fidelities for the response function calculation of KH.} (\textbf{A} to \textbf{B}) Fidelities between the raw experimental and exact system-qubit density matrices without (A) and with RC (B). (\textbf{C} to \textbf{D}) Fidelities between the purified experimental and exact system-qubit density matrices without (C) and with RC (D). Layout of the tiles in each panel is the same as in Fig.~\ref{fig:fidelity_matrix} in the main text. Similar to NaH, RC combined with McWeeny purification yields the highest system-qubit state fidelities.}
\label{fig:fidelity_matrix_kh}
\end{figure}

\begin{figure}[h]
\centering{
\phantomsubcaption\label{fig:chi00_norc_kh}
\phantomsubcaption\label{fig:chi00_rc_kh}
\phantomsubcaption\label{fig:chi01_norc_kh}
\phantomsubcaption\label{fig:chi01_rc_kh}
\includegraphics[width=0.62\textwidth]{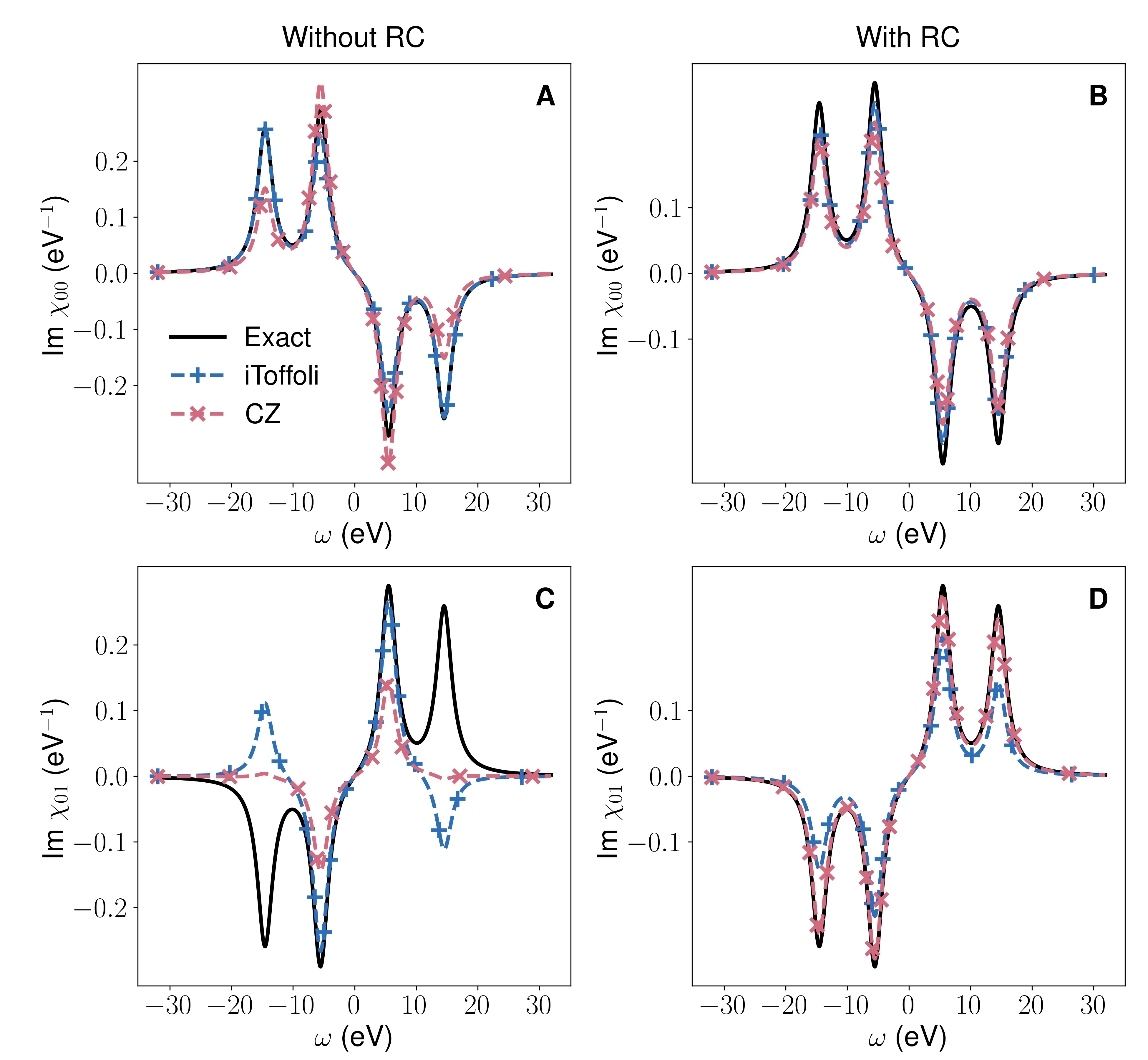}}
\caption{\textbf{Density-density response function of KH}. (A) $\Im\,\chi_{00}$ without RC. (B) $\Im\,\chi_{00}$ with RC. (C) $\Im\,\chi_{01}$ without RC. (D) $\Im\,\chi_{01}$ with RC.  All experimental results are processed with McWeeny purification. Similar to NaH, with RC and McWeeny purification, the results from iToffoli decompositions and CZ decompositions are comparable in reproducing the exact response function values.
}
\label{fig:response_function_kh}
\end{figure}

This section presents the system-qubit state fidelities and density-density response functions of the KH molecule. Figure \ref{fig:fidelity_matrix_kh} shows the system-qubit state fidelities in the response function calculations of KH. In the case of RC without purification, the average off-diagonal fidelity improved from 49.2\% in Fig.~\ref{fig:fidelity_matrix_norc_raw_kh} to 61.3\% in Fig.~\ref{fig:fidelity_matrix_rc_raw_kh}, whereas the average diagonal fidelity slightly decreased from 91.5\% in Fig.~\ref{fig:fidelity_matrix_norc_raw_kh} to 90.5\% in Fig.~\ref{fig:fidelity_matrix_rc_raw_kh}. Purification without RC shows a unified improvement across both diagonal and off-diagonal fidelities, which change to 97.1\% and 66.9\% from 49.2\% and 91.5\% respectively from Fig.~\ref{fig:fidelity_matrix_norc_raw_kh} to Fig.~\ref{fig:fidelity_matrix_norc_pur_kh}. The most significant improvement again comes from applying both purification and RC, with the average diagonal fidelity 99.8\% and average off-diagonal fidelity 95.9\% in Fig.~\ref{fig:fidelity_matrix_rc_pur_kh}.

Figure \ref{fig:response_function_kh} shows the density-density response functions of KH with all data postprocessed with McWeeny purification. As compared to the case for NaH, the results without RC only fail to reproduce the peaks or produce peaks with the wrong signs in Fig.~\ref{fig:chi01_norc_kh}, but reproduce all the peaks in Figs.~\ref{fig:chi00_rc_kh} and \ref{fig:chi01_rc_kh}. As for comparison between the iToffoli decomposition and CZ decomposition, the root-mean-squared deviations without RC are 0.0108 eV$^{-1}$ for iToffoli and 0.0316 eV$^{-1}$ for CZ in Fig.~\ref{fig:chi00_norc_kh}, while the deviations are 0.1007 eV$^{-1}$ for iToffoli and 0.0824 eV$^{-1}$ for CZ in Fig.~\ref{fig:chi01_norc_kh}. With RC, the root-mean-squared deviations are 0.0148 eV$^{-1}$ for iToffoli and 0.0225 eV$^{-1}$ for CZ in Fig.~\ref{fig:chi00_rc_kh}, and 0.0386 eV$^{-1}$ for iToffoli and 0.0064 eV$^{-1}$ for CZ in Fig.~\ref{fig:chi01_rc_kh}. In all cases with or without RC, the two decompositions show comparable results in the response functions.

\end{document}